# Reducing ion energy spread in hole-boring radiation pressure acceleration by using two-ion-species targets

S. M. Weng[1,*], M. Murakami[2], and Z. M. Sheng[1,3]

[1] Key Laboratory for Laser Plasmas, Department of Physics and Astronomy, Shanghai Jiao Tong University, Shanghai 200240, China

[2] Institute of Laser Engineering, Osaka University, Osaka 565-0871, Japan

[3] SUPA, Department of Physics, University of Strathclyde, Glasgow G4 0NG, UK

* E-mail: wengsuming@gmail.com; Tel: +86-21-34204629

## Abstract

The generation of fast ion beams in the hole-boring radiation pressure acceleration by intense laser pulses has been studied for targets with different ion components. We find that the oscillation of the longitudinal electric field for accelerating ions can be effectively suppressed by using a two-ion-species target, because fast ions from a two-ion-species target are distributed into more bunches and each bunch bears less charge. Consequently, the energy spread of ion beams generated in the hole-boring radiation pressure acceleration can be greatly reduced down to 3.7% according to our numerical simulation.

# 1. Introduction

Because of various advantageous features with a relatively low cost, laser-driven ion beams are of great interest for a wide range of potential applications: hadron therapy (Fourkal *et al.* 2007), proton imaging (Borghesi *et al.* 2002; Borghesi *et al.* 2010), neutron generation (Norreys *et al.* 1998), and fast ion ignition (Roth *et al.* 2001). Over the last decades, many interesting mechanisms have been proposed to accelerate ions for these applications (Daido *et al.* 2012; Yin *et al.* 2006; Badziak *et al.* 2010). However, there are still a few pivotal issues to be solved or improved for the large-scale application of laser-driven ion beams in practice. Among these issues, reducing the energy spread of generated fast ion beams particularly attracts intensive attention since quasi-monoenergetic ion beams are of critical benefit to some applications, such as hadron therapy and fast ion ignition. In order to generate quasi-monoenergetic ion beams, many novel designs of laser pulses and targets are proposed to reduce the energy spread of ion beams (Haberberger *et al.* 2012). In particular, many studies have showed that using two-ion-species targets can effectively stabilize ion acceleration and reduce the ion energy spread in the target normal sheath acceleration (TNSA) (Cui *et al.* 2013), as well as the "light sail" mode of radiation pressure acceleration (RPA) (Yu *et al.* 2010; Kar *et al.* 2012).

In this paper, we investigate the generation of quasi-monoenergetic ion beams in the "hole boring" mode of RPA by using two-ion-species targets. The hole-boring was firstly proposed as an approach to deliver the laser pulse closer to the core of a precompressed fusion fuel in the fast ignition scheme of inertial confinement fusion (Tabak *et al.* 1994; Kodama *et al.* 1996). However, the follow-up study indicated that the hole-boring is incompatible with the traditional fast electron ignition (Mulser & Schneider, 2004). On the other hand, a number of previous studies have demonstrated that the hot electron generation can be effectively suppressed under the irradiation of an intense circularly polarized pulse, and then fast ions can be efficiently generated from a thick dense target by the hole-boring RPA (Wilks *et al.* 1992; Macchi *et al.* 2005; Robinson *et al.* 2009a; Schlegel *et al.* 2009). Owing to the large number of ions at moderate energy generated, the hole-boring RPA provides a very promising approach to realizing fast ion ignition that requires an ion energy fluence as high as a few GJ cm$^{-2}$ at a few ten MeV/u (Badziak *et al.* 2011; Robinson, 2011; Naumova *et al.* 2009; Weng *et al.* 2014). However, a narrow-energy-spread ion beam is critically required for suppressing the time-of-flight spread and depositing the majority of ion beam energy into the hotspot in fast ion ignition (Hegelich *et al.* 2011). Based on particle-in-cell (PIC) simulations with different target components, we find that the energy spread of ion beam generated by the hole-boring RPA can be effectively reduced by using a two-ion-species target. In deuterium-tritium (DT) targets, the narrowest energy spread is roughly achieved in the target with deuterons and tritons at 1:1 mass ratio. This study may shed light on the appropriate target design for generating quasi-monoenergetic ion beams by the hole-boring RPA, which is particularly suitable for the application in the *in-situ* hole-boring fast ion ignition of the pre-compressed DT fuel (Naumova *et al.* 2009; Weng *et al.* 2014).

## 2. Theory of hole-boring acceleration

In a quasistationary piston model for the hole-boring RPA, the laser ponderomotive force pushes all electrons forward and then brings about a charge separation layer, while the ions left behind are pulled by the longitudinal electric field arising in this charge separation layer (Schlegel *et al.* 2009). Both the resulting hole-boring velocity $v_b$ and the accelerated ion energy $\varepsilon_i$ can be derived from the momentum flux balance in the boosted frame moving with the laser-plasma interface as (Macchi *et al.* 2005; Robinson *et al.* 2009a; Schlegel *et al.* 2009)

$$v_b = c\Pi/(1+\Pi), \quad \varepsilon_i = m_i c^2 2\Pi^2/(1+2\Pi), \tag{1}$$

with the speed of light in vacuum $c$, the ion mass $m_i$, and the hole-boring characteristic parameter $\Pi$ defined as

$$\Pi = (I/\rho c^3)^{1/2} = a(Zm_e n_c/2m_i n_e)^{1/2}, \tag{2}$$

where the dimensionless laser amplitude $a = (I\lambda^2/1.37\times 10^{18}\ \text{W cm}^{-2}\mu\text{m}^2)^{1/2}$, the critical density $n_c = m_e \varepsilon_0 \omega^2/e^2$, $\rho$ is the mass density, $Z$ is the ionic charge state, $m_e$ and $n_e$ are the electron mass and density, and $I$, $\lambda$ and $\omega$ are the laser intensity, wavelength and frequency, respectively.

However, the previous studies indicated that the longitudinal electric field for accelerating ions always oscillates due to the periodic ejection of ion bunches from the charge separation layer (Robinson *et al.* 2009a; Schlegel *et al.* 2009). On the one hand, the steady hole-boring acceleration with a well-controlled oscillating field in one-ion-species targets only takes place under the condition $a < n_e/n_c$, i.e., with a sufficiently high plasma density or a relatively low laser intensity (Schlegel *et al.* 2009). Otherwise, the oscillation of the electric field may be obviously enhanced and the energy distribution of fast ions will be broadened, which seriously limits its applications. On the other hand, Eq. (1) indicates that the sufficiently high laser intensity and low plasma density are required for generating ions at enough high energy. In the following PIC simulations, however, we will show that the oscillation of the longitudinal electric field in the hole-boring RPA can be well suppressed even under the condition

$a \geq n_e / n_c$ by using two-ion-species targets. Consequently, the energy spread of ion beams can be significantly reduced for accelerating the realization of its potential applications.

## 3. Hole-boring acceleration in two-ion-species targets

The employed PIC code is similar to that used in previous investigations (Weng *et al.* 2012a; Weng *et al.* 2012b; Weng *et al.* 2014). In all simulations, the pulse is assumed to be circularly polarized with the wavelength $\lambda = 1.06$ $\mu$m and thus the critical density $n_c = 9.92 \times 10^{20}$ cm$^{-3}$. The laser amplitude rapidly rises as $a\sin(\pi t/10\tau)$ in the first five cycles and then keeps constant with $a = 50$ and $\tau = 2\pi/\omega$. The simulation box is located at $-10\lambda \leq x \leq 70\lambda$ and simulations start from $t = -10\tau$, where the laser pulse is launched from the left boundary $x = -10\lambda$. The targets with different components are all located at $0 \leq x \leq 50\lambda$, and each one is initially cold with a homogeneous density. The spatial and temporal resolutions $\Delta x = \lambda/300$ and $\Delta t = \Delta x/c$ are used, and $20 n_s / n_c$ macroparticles per cell are initially allocated for the particle species with density $n_s$.

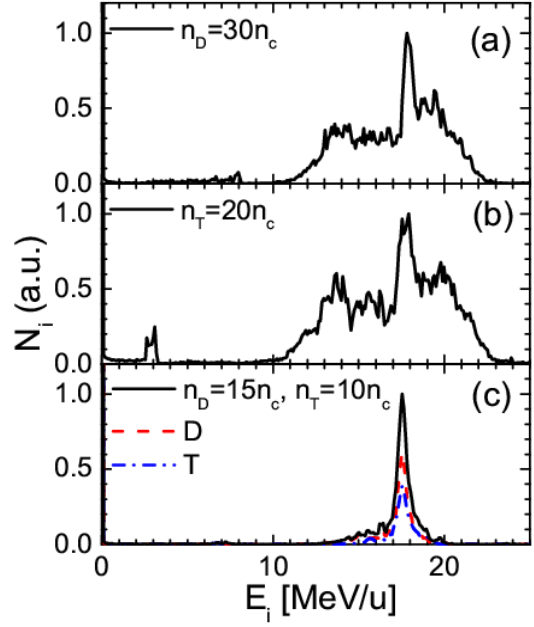

*Fig.1. (Color online) Ion energy spectra at $t = 100\tau$ from PIC simulations. The laser amplitude is $a = 50$ and targets are: (a) deuterium target with $n_e = n_D = 30 n_c$, (b) tritium target with $n_e = n_T = 20 n_c$, and (c) DT target with $n_D = 15 n_c$, $n_T = 10 n_c$, and $n_e = 25 n_c$. The respective spectra for deuterons (red dashed) and tritons (blue dash-dotted) from the DT target are also drawn in (c).*

Figure 1 shows the energy spectra of ions accelerated from three representative targets: (a) deuterium target, (b) tritium target, and (c) DT target with deuterons and tritons in 1:1 mass ratio. The electron number density varies in order to yield the same mass density $\rho = 0.1 \text{ g cm}^{-3}$ in all targets. Since the overall hole-boring dynamics are determined by the laser intensity and the target total mass density (Robinson *et al.* 2009b), it is reasonable that ions are roughly accelerated to the same velocity and hence the ion spectra from three targets are all peaked at the same energy per nucleon (17.55 MeV/u) as predicted by Eq. (1). For the applied laser pulse $a = 50$, it is clear that the condition $a < n_e / n_c$ fails for all targets.

Therefore, it is expected that the resulting energy spectra should be quite broad. Indeed, this is true in the target with either pure deuterons or pure tritons. As shown in Fig. 1(a) and (b), the accelerated deuterons or tritons are broadly distributed over the energy spectrum with the respective energy spreads $\Delta E/E$=13.2% or 38.9%, here $\Delta E$ is the full width at the half maximum (FWHM) and $E$ is the peak energy of the spectrum. However, the energy spectrum from the DT target surprisingly displays a favorable quality of being quasi-monoenergetic with a narrow spread 3.7% in Fig. 1(c). It is important to note that the principle of reducing the energy spread in the hole-boring RPA by using two-ion-species targets is essentially different from these in the TNSA or light-sail RPA. In the TNSA or light-sail RPA, the lighter ions such as protons are quickly separated from the heavier ions, then the heave ion layer acts as a substrate to protect the lighter ions from the instabilities and thus provide a more stable accelerating field for generating quasi-monoenergetic lighter ions (Yu *et al.* 2010; Cui *et al.* 2013). Usually, the energy spread of the lighter ions monotonously decreases with its ratio and the quality of heavier ion beams is sacrificed in the TNSA or light-sail RPA using two-ion-species targets. In the hole-boring RPA with two-ion-species targets, however, it is clear that the respective spectra of lighter ions (deuterons) and heavier ions (tritons) are very similar as shown in Fig. 1(c). They are almost peaked at the same energy per nucleon with narrow spreads 3.2% and 3.7%, respectively.

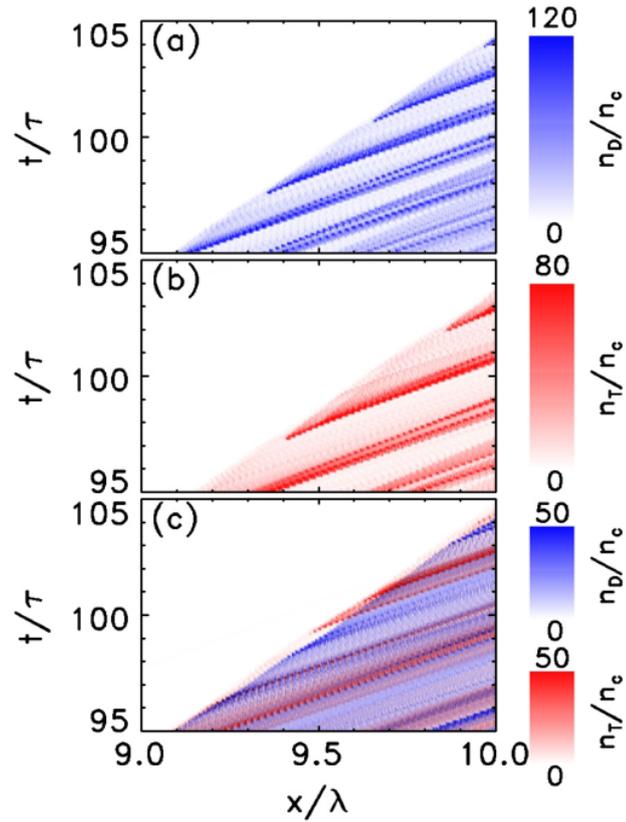

*Fig.2. (Color online) Time evolution of the density of only fast ions with energy $\geq 10$ MeV/u in different targets, the laser and target parameters are the same as in figure 1.*

In order to understand the distinct ion energy spectra from different targets, it is essentially helpful to study the dynamics of accelerated ions. The time evolution of the density of only fast ions with energy $\geq 10$ MeV are compared in Fig. 2. As previously predicted, the accelerated ions by the hole-boring RPA will leave the charge separation layer in bunches rather than as a constant current because the laser pulse does not act as an ideal piston (Robinson *et al.* 2009a; Schlegel *et al.* 2009). This prediction is well illustrated by the streaks for fast deuterons or tritons in Fig. 2(a) or (b), respectively. It is clearly shown that the accelerated ions from one-ion-species targets are separated into bunches. The time interval among these ion bunches keeps nearly constant in the acceleration, and its value is inversely proportional to the ion plasma frequency (Schlegel *et al.* 2009)

$$\Delta t \propto \frac{1}{\omega_{pi}} = \sqrt{\frac{m_i n_c}{m_e n_e}} \frac{1}{\omega}, \tag{3}$$

While the accelerated deuterons and tritons from the DT target will depart from the charge separation layer in respective bunches because of their different plasma frequencies. Further, the time interval among ion bunches does no longer keep constant in the acceleration as indicated by the irregular bands of color in Fig. 2(c). As a result, the accelerated deuterons and tritons from the DT target are distributed into more bunches and each bunch bears less charge than these from the deuterium or tritium target.

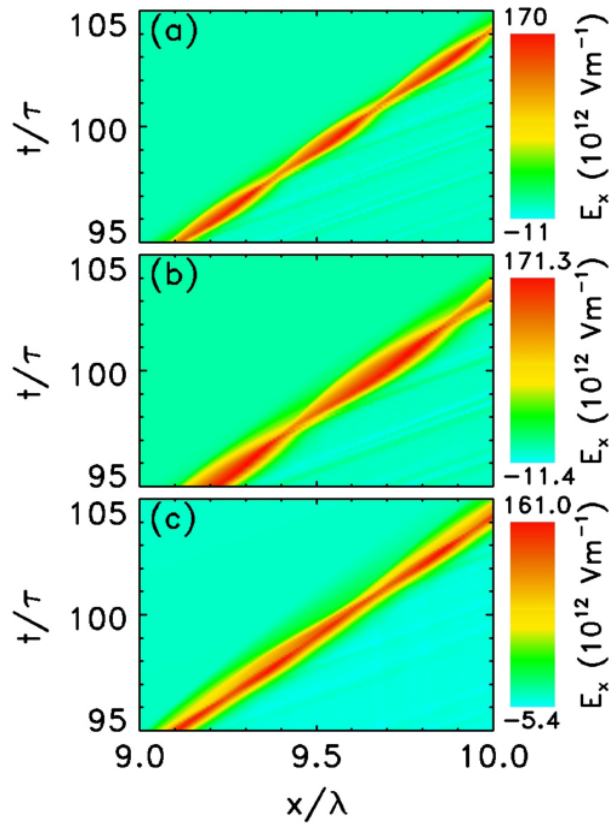

*Fig.3. (Color online) Time evolution of the longitudinal electric field $E_x$ in different targets, the laser and target parameters are the same as in figure 1.*

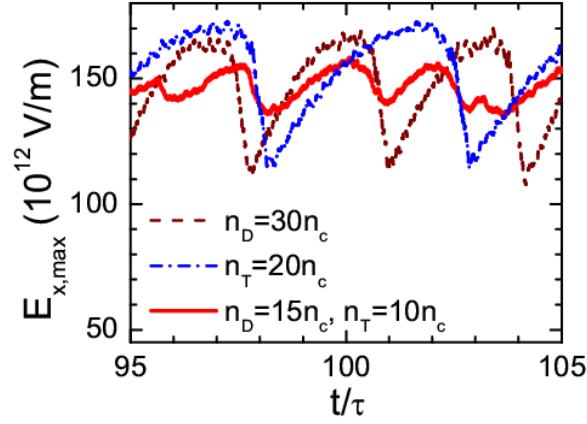

*Fig.4. (Color online) Time evolution of the magnitude of the longitudinal electric field $E_{x,\max}$ in three different targets, the laser and target parameters are the same as in figure 1.*

Since the accelerated ions leave the charge separation layer in bunches, the total charge number in this layer will fluctuate. Consequently, the longitudinal electric field $E_x$ oscillates as shown in Fig. 3, in which the saffron yellow bands also roughly indicate the range of the charge separation layer. For the deuterium target, the saffron yellow band in Fig. 3(a) contracts at the moments 95, 98, 101, $104\tau$, while the ejections of fast ion bunches also happen at the corresponding moments as shown in Fig. 2(a). The synchronization between the ejection of fast ion bunches and the contraction of charge separation layer is similarly identified by Fig. 2(b) and Fig. 3(b) for the tritium target. For the DT target, we have found that the time evolution of the longitudinal electric field becomes smoother within a uniform charge separation layer as shown in Fig. 3(c). The time evolution of the magnitude of the longitudinal electric field are compared among three targets in Fig. 4. It is shown that the relative oscillation of the field magnitude $\Delta E_{x,\max} / E_{x,\max}$ are roughly 40%, 42%, and 13% for the deuterium, tritium, and DT targets, respectively. This clearly indicates that the oscillation of the accelerating field in the hole-boring RPA can be effectively suppressed by using a DT target.

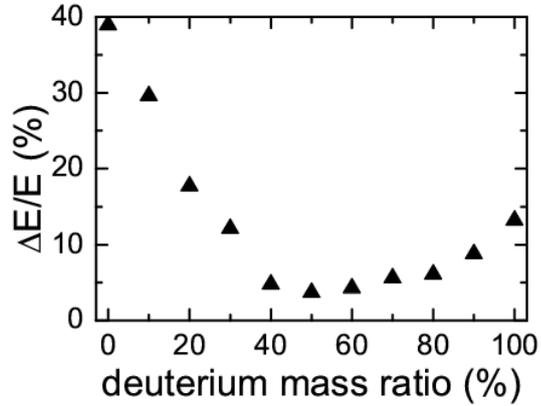

*Fig.5. Ion energy spread $\Delta E/E$ as a function of the deuterium mass ratio in the composite DT targets at $t=100\tau$ from PIC simulations. The laser parameters are the same as in figure 1, and the total mass density of composite targets is fixed at $\rho=0.1 \text{ g cm}^{-3}$.*

Finally, the effect of the mass ratio between deuterons and tritons in the composite targets are studied with the fixed total mass density $\rho=0.1 \text{ g cm}^{-3}$. With the increasing mass ratio of deuterons, the energy spread of ion spectrum decreases dramatically at first, then it rebounds slowly. The narrower energy spread from the pure deuterium target than the pure tritium target could be attributed to its higher electron density. Overall, it is clear that the narrowest energy spread is achieved roughly at the deuterium mass ratio of 50%.

## 4. Conclusion

In conclusion, we have found that the energy spread of ion beams generated by the hole-boring RPA can be effectively reduced by using two-ion-species targets. In contrast to the one-ion-species target, the lighter and heavier ions in the two-ion-species target depart from the charge separation layer in more, but smaller, bunches. Consequently, the fluctuation in the charge number within the charge separation layer is weakened and so does the oscillation of the longitudinal electric field. Therefore, the energy spread of fast ions can be dramatically reduced. Unlike these by the TNSA or the light-sail RPA, the energy spread of fast heavier ions generated by the hole-boring RPA in a two-ion-species target is nearly as narrow as that of lighter ions, which may be beneficial to enhancing the energy conversion efficiency of laser pulse into useful fast ions.

## ACKNOWLEDGEMENTS

The authors are grateful for the use of PI cluster at Shanghai Jiao Tong University. SMW acknowledges a fellowship provided by the Japan Society for the Promotion of Science (JSPS) during his stay in Osaka University, when part of this work was done. This work was supported in part by the National Basic Research Program of China (Grant No. 2013CBA01504), the National Natural Science Foundation of China (Grant Nos. 11405108, 11129503 and 11374210), and the MOST international collaboration project (Grant No. 2014DFG02330).

**References:**

1. Badziak, J., Jabłoński, S. & Parys, P. *et al.* (2010). Production of high-intensity proton fluxes by a 2ω Nd: glass laser beam. *Laser Part. Beams* **28**, 575-583.

2. Badziak, J., Mishra, G., Gupta, N. K. & Holkundkar, A. R. (2011). Generation of ultraintense proton beams by multi-ps circularly polarized laser pulses for fast ignition-related applications. *Phys. Plasmas* **18**, 053108.

3. Borghesi, M., Campbell, D. H. & Schiavi, A. *et al.* (2002). Electric field detection in laserplasma interaction experiments via the proton imaging technique. *Phys. Plasmas* **9**, 2214-2220.

4. Borghesi, M., Sarri, G. & Cecchetti, C. A. *et al.* (2010). Progress in proton radiography for diagnosis of ICF-relevant plasmas. *Laser Part. Beams* **28**, 277-284.

5. Cui, Y. Q., Wang, W. M., Sheng, Z. M., Li, Y. T. & Zhang, J. (2013). Quasimonoenergetic proton bunches generation from doped foil targets irradiated by intense lasers. *Phys. Plasmas* **20**, 024502.

6. Daido, H., Nishiuchi, M. & Pirozhkov, A. S. (2012). Review of laser-driven ion sources and their applications. *Rep. Prog. Phys.* **75**, 056401 and references therein.

7. Fourkal, E., Velchev, I., Fan, J., Luo, W. & Ma, C.-M. (2007). Energy optimization procedure for treatment planning with laser-accelerated protons. *Med. Phys.* **34**, 577-584.

8. Haberberger, D., Tochitsky, S. & Fiuza, F. (2012). Collisionless shocks in laser-produced plasma generate monoenergetic high-energy proton beams. *Nature Phys.* **8**, 95-99.

9.  Hegelich, B. M., Jung, D. & Albright, B. J. *et al.* (2011). Experimental demonstration of particle energy, conversion efficiency and spectral shape required for ion-based fast ignition. *Nucl. Fusion* **51**, 083011.

10. Kar, S., Kakolee, K. F. & Qiao, B. *et al.* (2012). Ion Acceleration in Multispecies Targets Driven by Intense Laser Radiation Pressure. *Phys. Rev. Lett.* **109**, 185006.

11. Kodama, R., Takahashi, K., & Tanaka, K. A. *et al.* (1996). Study of Laser-Hole Boring into Overdense Plasmas. *Phys. Rev. Lett.* **77**, 4906-4909.

12. Macchi, A., Cattani, F., Liseykina, T. V. & Cornolti, F. (2005). Laser Acceleration of Ion Bunches at the Front Surface of Overdense Plasmas. *Phys. Rev. Lett.* **94**, 165003.

13. Mulser, P., & Schneider, R. (2004). On the inefficiency of hole boring in fast ignition. *Laser Part. Beams* **22**, 157-162.

14. Naumova, N., Schlegel, T. & Tikhonchuk, V. T. *et al.* (2009). Hole Boring in a DT Pellet and Fast-Ion Ignition with Ultraintense Laser Pulses. *Phys. Rev. Lett.* **102**, 025002.

15. Norreys, P. A., Fews, A. P. & Beg, F. N. *et al.* (1998). Neutron production from picosecond laser irradiation of deuterated targets at intensities of $10^{19}$ W cm$^{-2}$. *Plasma Phys. Control. Fusion* **40**, 175-182.

16. Robinson, A. P. L., Gibbon, P. & Zepf, M. *et al.* (2009). Relativistically correct hole-boring and ion acceleration by circularly polarized laser pulses. *Plasma Phys. Control. Fusion* **51**, 024004.

17. Robinson, A. P. L., Kwon, D. H. & Lancaster, K. (2009). Hole-boring radiation pressure acceleration with two ion species. *Plasma Phys. Control. Fusion* **51**, 095006.

18. Robinson, A. P. L. (2011). Production of high energy protons with hole-boring radiation pressure acceleration. *Phys. Plasmas* **18**, 056701.

19. Roth, M., Cowan, T. E. & Key, M. H. *et al.* (2001). Fast Ignition by Intense Laser-Accelerated Proton Beams. *Phys. Rev. Lett.* **86**, 436-439.

20. Schlegel, T., Naumova, N. & Tikhonchuk, V. T. *et al.* (2009). Relativistic laser piston model: Ponderomotive ion acceleration in dense plasmas using ultraintense laser pulses. *Phys. Plasmas* **16**, 083103.

21. Tabak, M., Hammer, J. & Glinsky, M. E. *et al.* (1994). Ignition and high gain with ultrapowerful lasers. *Phys. Plasmas* **1**, 1626-1634.

22. Weng, S. M., Mulser, P. & Sheng, Z. M. (2012). Relativistic critical density increase and relaxation and high-power pulse propagation. *Phys. Plasmas* **19**, 022705.

23. Weng, S. M., Murakami, M., Mulser, P. & Sheng, Z. M. (2012). Ultra-intense laser pulse propagation in plasmas: from classic hole-boring to incomplete hole-boring with relativistic transparency. *New J. Phys.* **14**, 063026.

24. Weng, S. M., Murakami, M. & Azechi, H. *et al.* (2014). Quasi-monoenergetic ion generation by hole-boring radiation pressure acceleration in inhomogeneous plasmas using tailored laser pulses. *Phys. Plasmas* **21**, 012705.

25. Wilks, S. C., Kruer, W. L., Tabak, M. & Langdon, A. B. (1992). Absorption of ultra-intense laser pulses. *Phys. Rev. Lett.* **69**, 1383-1386.

26. Yin, L., Albright, B. J., Hegelich, B. M. & Fernández, J. C. (2006). GeV laser ion acceleration from ultrathin targets: The laser break-out afterburner. *Laser Part. Beams* **24**, 291-298.

27. Yu, T. P., Pukhov, A., Shvets, G. & Chen, M. (2010). Stable Laser-Driven Proton Beam Acceleration from a Two-Ion-Species Ultrathin Foil. *Phys. Rev. Lett.* **105**, 065002.